# Effect of toroidal moment on a macroscopic self-organization of electrons in the quantum Hall regime

*S.A.Emelyanov*[*]

*Division of Solid State Electronics, A.F.Ioffe Institute, 194021 St.Petersburg, Russia*

**Abstract**

We have studied CR lineshape of terahertz-light-induced current in InAs quantum wells in tilted quantizing magnetic fields. We have observed dramatic modification of the lineshape with increasing of in-plane component of magnetic field as well as with increasing of transverse built-in electric field in the well. Scenario of the modification shows that the energy spectrum asymmetry is determined by so-called toroidal moment of the system and is a function of Landau quantum number. Macroscopic self-organization of electrons under the conditions of quantum Hall effect has also been directly demonstrated in both linear and saturation regimes of the light absorption.

*PACS*: 73.43.Fj, 73.22.Gk

*Key words*: A. Quantum wells; D. Electronic band structure; D. Photovoltaics

## 1. Introduction

Toroidal moments together with electric and magnetic ones are the three independent families of electromagnetic multipoles which should be taken into account to describe completely an electronic system with an arbitrary current and charge distribution there (for a review see [1] and refs. therein). The concept of toroidal moments has also been discussed in the context of possibility of a super-diamagnetism in non-superconducting systems [2]. The idea that an in-plane external magnetic field may give rise to a nonzero toroidal moment in an asymmetric two-dimensional (2D) system was first advanced in [3, 4]. Using simplified theoretical models it has been shown that the energy spectrum of such systems should be asymmetrical in *k*-space so that the asymmetry is characterized by a polar vector:

$$\vec{T} = \vec{B} \times \vec{E}, \tag{1}$$

where $\vec{B}$ is external magnetic field, $\vec{E}$ is potential gradient perpendicular to the well plane which is known as a "built-in" electric field. Vector $\vec{T}$ is just the system's toroidal moment.

The interest to the problem of toroidal moments has been stimulated by recent experiments in which a some sort of self-organization of electrons has been observed just in a toroidal-moment-possessing 2D system which is under the conditions of the quantum Hall effect [5]. The self-organization has been shown to occur when the Landau level degeneracy is lifted by a combined effect of both built-in electric field and tilted quantizing magnetic field that leads to the breaking of space-inversion and time-reversal symmetry of the system. The intriguing feature is while 2D electron gas is clearly a microscopic system, the self-organization has been demonstrated to be of a macroscopic character. In the experiments, it manifests itself in appearance of a set of parallel oncoming electric currents under homogeneous photoexcitation of 2D system in presence of tilted quantizing magnetic field. To explain qualitatively the effect observed, the models of Refs. [3, 4] have been applied. Although many specific features of the effect are described well by these models, the number of problems appear. First, the energy spectrum of the system is still not calculated either analytically or numerically. Second, most principle, problem is the models imply the existence of spontaneous non-dissipative oncoming currents in the system which are essentially one-dimensional. Moreover, in the experiment the system behaves as if these currents are truly one-dimensional. On the other hand, it is clear from general considerations that such currents may occur only in an endless system. The existence of them in the obviously certain system seems to be quite paradoxical and requires an adequate understanding.

---

[*] E-mail: sergey.emelyanov@mail.ioffe.ru

In present paper, we study the behaviour of 2D system in the regime of macroscopic self-organization of the electron gas by analysis of CR lineshape of terahertz-light-induced currents in InAs quantum wells in presence of both an external tilted quantizing magnetic field and built-in electric field in the well. We demonstrate that the energy spectrum asymmetry is determined by the so-called toroidal moment of the system as well as by the Landau quantum number The results have been found to be consistent with the idea of transformation of the Landau levels into the bands asymmetrical in $k$-space, in which the electrons are macroscopically self-organized and provide 1D spontaneous non-dissipative currents in the system.

## 2. Experimental

The samples used in the experiments are MBE-grown InAs/GaSb structures containing a 15 nm wide InAs single quantum well surrounded by AlSb barriers to avoid hybridization-related effects. The samples of three types have been utilized. The type-one samples are not intentionally doped ones. The lower AlSb barrier is as thick as 0.1 μm while the upper one is only 3 nm wide covered by 20 nm GaSb protecting layer. These samples have also been capped by a 5 nm wide InAs layer to minimise surface potential effect on the well [6]. Type-two samples are those in which two tellurium δ-doping layers are incorporated: 15 nm below and 15 nm above the well. Two AlSb barriers are 40 nm wide and the upper one is capped by 6 nm wide GaSb protecting layer. Finally, one-side-doped samples have been used (type-three ones) which are similar to those of type-two but here the only one δ-doping layer was incorporated 15 nm below the well. The electron sheet density in the samples ranged from 1.6 to $2.0 \times 10^{12}$ cm$^{-2}$ (mobility about $10^5$ cm$^2$/Vs). Relative value of built-in electric field in the samples was estimated through the non-resonant photo-galvanic effect which occurs in presence of in-plane non-quantizing magnetic field. As expected, the highest field is in the type-three samples while the lowest field is in the type-one samples. In all samples the built-in field is supposed to be pointed toward the capping layer. The samples (typically $8 \times 8$ mm) are supplied by either solid or dashed indium ohmic contacts. Geometry of the contacting is shown in Fig. 1(a, b).

Terahertz-radiation-induced in-plane currents in unbiased samples were studied in presence of tilted quantizing magnetic fields under cyclotron resonance (CR) conditions. As a source of terahertz radiation we have used pulsed gas laser optically pumped by $CO_2$ laser. The active medium was either heavy water ($D_2O$) or ammonia ($NH_3$). The wavelength was 385 μm ($\hbar\omega = 3.2$ meV) and 90.6 μm ($\hbar\omega = 13.7$ meV), respectively. In the former case pulse duration was 80 ns, maximum radiation intensity $I_m \approx 40$ W/cm$^2$ while in the later case 40 ns and 200 W/cm$^2$, respectively. In both cases the laser radiation was linearly polarized. The in-plane current pulses were detected by a high-speed storage oscilloscope through the voltage drop on a 50 Ohm load resistor in a short-circuit regime. The incident radiation was normal to the sample surface. The experiments were performed at $T = 1.9$ K. Geometry of the experiments is shown in Fig. 1(a, b).

## 3. Results and discussion.

Fig. 2 shows light-induced current in type-one samples vs quantizing component of magnetic field ($B_Z$) at $\lambda = 385$ μm in the geometry of Fig. 1(a) after the subtracting of a non-resonant component of the current. The measurements have been done at three different $B_X$. It is seen that the increasing of $B_X$ results in a dramatic modification of the resonance lineshape. At low $B_X$, the lineshape is reminiscent the Lorentzian one while at higher $B_X$ it tends to become a bipolar and looks like a sum of two resonances of different polarity shifted from each other on $B_Z$ scale. Position of the resonances on the magnetic field scale clearly shows that we deal with a CR-related effect since the deducted effective mass ($m^* \approx 0.04 m_0$) is consistent with the commonly used value for such a system. Broadening of the Lorentzian-like curve (curve 1) is not higher than that usually observed in transmission spectra so that means the effect of saturation of the laser light absorption is supposed to be of a minor importance here. To compare the effect of in-plane magnetic field on CR lineshape with that of built-in electric field, we have studied the behaviour of light-induced current in the vicinity of CR point at $\lambda = 90.6\,\mu m$ for all three types of the samples. In these experiments, quantizing component of magnetic field is varied while the in-plane component is constant. The results are shown in Fig. 3. It is clearly seen that scenario of modification of CR lineshape is qualitatively the same as that in Fig. 2: at low built-in field CR lineshape is reminiscent the Lorentzian one but at higher fields it tends to become a bipolar and, once more, looks like a sum of two resonances of different polarity shifted from each other on $B_Z$ scale. Therefore, one can suppose that CR lineshape is determined by the product of $B_X$ and $E$. According to Eq. 1, this product is just the system's toroidal moment.

Consider now CR lineshape in more details. In general, a bipolar lineshape of light-induced current resonance is a characteristic feature of current resonances caused by vertical optical transitions between two electronic bands. Indeed, such a lineshape has been observed under resonant absorption between Landau subbands in a bulk system [7] as well as between quantum subbands in a 2D system [8]. In both cases the bands are fully symmetrical in *k*-space while an asymmetry was provided by excitation: the wave vector of exciting light was parallel to the resonant current. However, the same result can also be obtained under symmetrical photo-excitation but in presence of an inner system asymmetry. More specifically, if the bands are shifted from each other in *k*-space, then, at fixed light quantum, the resonant conditions will be fulfilled at different magnetic fields for two optical transitions symmetrical with respect to the lower band minimum. Because of the opposition of electrons' velocities for these transitions, one would expect a bipolar lineshape of the resonance. This picture is illustrated schematically in Fig. 4.

Rigorously, at least two more questions should also be discussed in the context of analysis of CR lineshape. The first one is whether the Landau level broadening is higher than the mutual shift of the bands or not? Evolution of the lineshape with increasing of $B_X$ allows one to suppose that the Lorentzian-like lineshape is attributed to the former case while the bipolar one is attributed to the latter. The second one is whether the bands are entirely the same or their shape is also a function of the Landau quantum number? To our opinion, CR lineshape of curve 2 in Figs. 2 and 3 allows one to suppose that the latter case is much more probable. Note also, that in Ref. [5] a weak declination of CR lineshape from the Lorentzian-like in photocurrent spectra has been interpreted in terms of the effect of built-in electric field on the electrons possessing nonzero electric dipole moment (Stark effect). Here we show that to explain such a lineshape, it is quite enough to suppose both position and shape of the bands to be a function of Landau quantum number. To clarify the question regarding the role of Stark effect as well as regarding the positions and the shape of the bands, self-consistent calculations of the energy spectrum is clearly required.

The crucial point in the context of presented results is whether the light-induced current along *Y* axis monitored in the experiments are truly consists of a set of spatially separated oncoming parallel currents as it follows from the results of Ref. [5]? To clarify this point, we have carried out the experiments on the samples supplied by three pairs of ohmic contacts as it is shown in Fig. 1(b). The contacts were arranged symmetrically with respect to the center of the sample. The length of each contact is 1.5 mm as well as the distances between them. All of them are at least 1 mm away from the sample border to avoid any border-related effects. The not-intentionally doped samples have been utilized. The experiments were performed at $\lambda = 90.6\,\mu m$ and at low tilt angles when CR lineshape in the geometry of Fig. 1(a) is a Lorentzian-like. Behaviour of the current through each pair of contacts vs quantizing component of magnetic field is shown in Fig. 5(a). It is clearly seen that despite of homogeneity of the photo-excitation, the currents through the contacts 1-2, 3-4 and 5-6 differ drastically from each other so that they may flow even in opposite directions simultaneously. This signifies that the electrons in the system are truly macroscopically self-organized because, otherwise, no reasons for strongly different photo-responses from the contacts oriented along the same axis. The origin of such a self-organization has been discussed in [5] where it has been supposed that the electrons on Landau levels are lined up in accordance with the following familiar formula known from the quantum Hall effect (QHE):

$$x_0 = -k_y r^2, \qquad (2)$$

where $x_0$ is coordinate of the center of electron's cyclotron orbit along *X* axis, $k_y$ is electron's wave vector along *Y* axis and $r$ is the magnetic length. Under conventional QHE, when external magnetic field is exactly perpendicular to the well plane, this formula plays obviously the role of a mathematical abstraction and can not lead to any real self-organization of electrons because of the axial symmetry of the system in the well plane. However, if we assume that the Landau level degeneracy can be lifted by a combined effect of built-in electric field and *X*-component of quantizing magnetic field while Eq. (2) remains, then we obtain immediately the following self-organization of electrons: their velocity along *Y* axis has become in a one-to-one correspondence with *X*-coordinate of the center of their cyclotron orbit. The latter parameter is known to be a macroscopic one since it ranges from $-L/2$ to $L/2$ (*L* is the sample length along *X* axis). Thus, just this parameter can be responsible for the macroscopic character of the self-organization. Further, in such a system any resonant vertical optical transition existing in some point of *k*-space may occur only in a corresponding point of real space. This may result in a set of spatially separated resonant light-induced currents along *Y* axis, i.e. just what we observe in the experiment.

As it is seen from Figs. 3 and 5(a), the width of Lorentzian-like CR lines at λ = 90.6 μm is sufficiently wider than that observed at λ = 385 μm (Fig. 2). This means at λ = 90.6 μm we are in a saturation regime that is consistent also with transmission spectra of Ref. [9]. To clarify the role of such a non-linearity, we have studied distribution of the currents over the sample at the same experimental conditions as in Fig. 5(a) but at one order lower laser intensity. The results are shown in Fig. 5(b). One can see that the characteristic features of the effect remains to be as pronounced as in the case of high excitation. This means that, most likely, nonlinear regime is not a crucial point for the self-organization effect. Taken into account that the effect occurs also at λ = 385 μm when we are obviously in a linear regime, one can conclude that the self-organization of electrons occurs in both linear and saturation regimes but in the latter case the current distribution over the sample may be a function of the laser intensity. Indeed, comparison of Figs. 5(a) and 5(b) shows that decreasing of laser intensity does not result in reverse of the sign of resonant current through the both 1-2 and 5-6 contacts. In contrast, the current through the contacts 3-4 reverses its sign and, moreover, its lineshape is no more a Lorentzian-like but reminiscent rather that of curve 2 in Figs. 2 and 3. Qualitatively, this result is also consistent with proposed physical picture. Indeed, contribution of each optical transition to the total current is determined primarily by electrons' velocity at corresponding point of *k*-space. In saturation regime, variation of the laser intensity should result in a redistribution of the relative intensities of the optical transitions. Hence, if the electrons' velocity is truly correlated with their position in real space, then one would expect a redistribution of light-induced currents over the sample. Moreover, one would expect the strongest redistribution in those region of the sample for which the corresponding fragment of the energy spectrum is farthest from monotonic function. Our results has shown that such a region is close to the center of the sample (the point $L = 0$) and, consequently, close to the point $k = 0$ in *k*-space. A more detailed map of the light-induced current distribution in the self-organization regime will be the matter of further investigation.

**Summary**

CR lineshape of terahertz-light-induced currents in asymmetric InAs quantum wells has been studied in presence of tilted quantizing magnetic fields under the conditions of quantum Hall effect. Dramatic modification of CR lineshape has been observed with increasing of in-plane component of magnetic field as well as with increasing of transverse built-in electric field in the well. Scenario of both modifications is found to be the same that allows one to suppose the energy spectrum asymmetry to be determined by so-called toroidal moment of the system as well as by the Landau quantum number. The existence of a macroscopic self-organization of electrons under the conditions of QHE has been directly demonstrated in both linear and saturation regimes of the light absorption through the observation of spatially separated parallel oncoming in-plane currents under homogeneous photo-excitation. In the saturation regime, the effect of redistribution of light-induced currents over the samples has been observed.

**Acknowledgements**

The MBE samples were kindly provided by B.Ya. Meltser and S.V. Ivanov (Ioffe Institute, St.Petersburg) as well as by Yu.G.Sadofyev (State Radio-Engineering Academy, Ryazan).

**Figures**

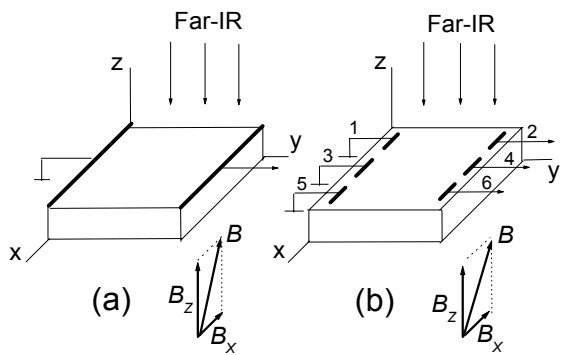

Fig. 1. Geometry of the experiments: (a) – solid contacting; (b) – dashed contacting.

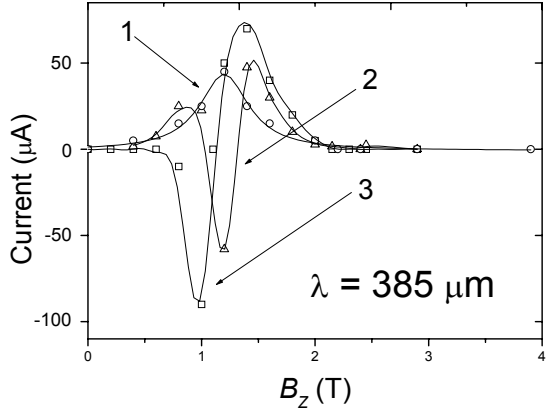

Fig. 2. Light-induced current as a function of quantizing component of magnetic field at fixed in-plane component after the subtracting of a non-resonant signal: 1. – $B_X = 0.15$ T; 2. – $B_X = 0.5$ T; 3. – $B_X = 1$ T. The type-one samples are used in the geometry of Fig. 1(a). Lattice temperature is 1.9 K.

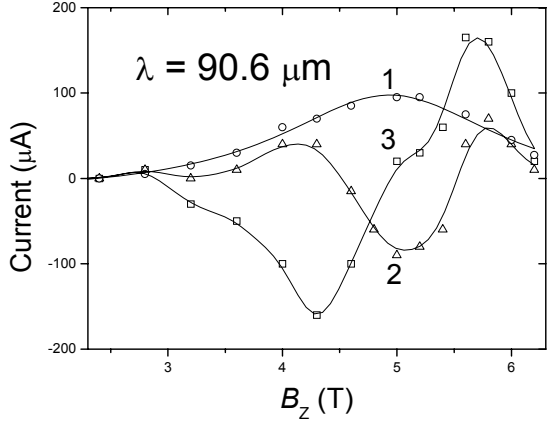

Fig. 3. Light-induced current as a function of quantizing component of magnetic field at fixed in-plane component ($B_X = 0.6$ T) for the three types of the samples possessing of different absolute value of built-in electric field: $E_1 < E_2 < E_3$. Geometry of the experiment is shown in Fig. 1(a). Lattice temperature is 1.9 K.

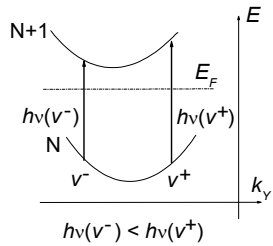

Fig. 4. Schematic illustration of a bipolar CR lineshape for the light-induced current along $Y$ axis. The resonant optical transitions occur between two bands shifted slightly from each other along $k_Y$.

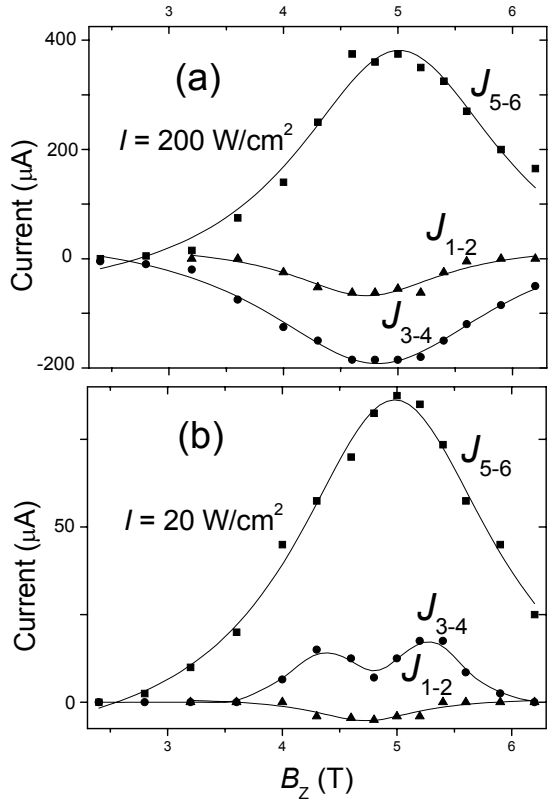

Fig. 5. Light-induced current through each of three pairs of ohmic contacts oriented along *Y* axis as a function of quantizing component of magnetic field ( $B_X = 0.6$ T). Geometry of the experiment is shown in Fig. 1(b). Lattice temperature is 1.9 K. The type-one samples are used. The laser wavelength is 90.6 μm while the intensity is varied: (a) - $I = 200$ W/cm$^2$; (b) - $I = 20$ W/cm$^2$.